# Quantum Generalized Linear Models

Colleen M. Farrelly
Co-Founder
Quantopo, LLC
cfarrelly@med.miami.edu

Srikanth Namuduri
Researcher
Florida International University
snamudur@fiu.edu

Uchenna Chukwu
Co-Founder
Quantopo, LLC
uchenna.chukwu@quantopoinc.com

## Abstract

Generalized linear models (GLM) are link function based statistical models. Many supervised learning algorithms are extensions of GLMs and have link functions built into the algorithm to model different outcome distributions. There are 2 major drawbacks when using this approach in applications using real world datasets. One is that none of the link functions available in the popular packages is a good fit for the data. Second, it is computationally inefficient and impractical to test all the possible distributions to find the optimum one. In addition, many GLMs and their machine learning extensions struggle on problems of overdispersion in Tweedie distributions.

In this paper we propose a quantum extension to GLM that overcomes these drawbacks. A quantum gate with non-Gaussian transformation can be used to continuously deform the outcome distribution from known results. In doing so, we eliminate the need for a link function. Further, by using an algorithm that superposes all possible distributions to collapse to fit a dataset, we optimize the model in a computationally efficient way. We provide an initial proof-of-concept by testing this approach on both a simulation of overdispersed data and then on a benchmark dataset, which is quite overdispersed, and achieved state of the art results. This is a game changer in several applied fields, such as part failure modeling, medical research, actuarial science, finance and many other fields where Tweedie regression and overdispersion are ubiquitous.

## Background
### Generalized Linear Models

Generalized linear models (GLM) are an extension and a generalization of simple linear regression. Whereas linear regression assumes a Gaussian distribution of the dependent variable, GLMs do not have such a limitation. Instead, they use a link function to model the relationship between the dependent variable and the linear combination of independent variables. [22, 24]

$$g(y) = \beta_0 + \beta_1 x_1 + \beta_2 x_2 + \cdots + \beta_n x_n$$

where g(y) is the link function.

This equation can be expressed in a matrix form by rearranging the terms to yield:

$$E(Y) = \mu = g^{-1}(X\beta)$$

with

$$\text{Var}(Y) = Var(\mu) = Var(g^{-1}(\boldsymbol{X}\beta))$$

Here Y is a vector of outcome values, μ is the mean of Y, **X** is the matrix of predictor values, g is a link function (such as the log function), and β is a vector of predictor weights in the regression equation. [22]

When expressed this way, the link function achieves something significant - it transforms the distribution of the dependent or outcome variable to a normal distribution in order to fit a linear model, yielding a generalized linear model. [22]

**Applications of GLM**

GLMs have many statistical extensions, including generalized estimating equations for longitudinal data modeling, generalized linear mixed models for longitudinal data with random effects, generalized additive models (where the predictor vectors can be transformed within the model), a survival data modeling with Cox regression and Weibull-based regression. **[22]**

GLMs themselves are ubiquitous in part failure modeling, medical research, actuarial science, and many other problems. Examples are modeling likelihood of insurance claims and expected payout (worldwide, a $5 trillion industry), understanding risk behavior in medical research (daily heroin usage, sexual partners within prior month…), modeling expected failure rates and associated conditions for airplane parts or machine parts within a manufacturing plant (~$4 trillion industry in the USA alone), and modeling expected natural disaster impacts and precipitating factors related to impact extent, among many others. [4, 6, 7, 9, 10, 11, 12, 15, 20, 23, 27]

**Tweedie Regression (13, 14, 21, 22)**

The error distributions in GLMs are modeled using exponential dispersion models, with Tweedie distributions as a general case of such exponential dispersion. Tweedie distributions have very useful geometric properties, and many common distributions of the exponential family converge to Tweedie distributions and can be formulated through Tweedie distributions, which are formally defined as:

$$E(Y) = \mu$$

$$Var(Y) = \varphi\mu^{\xi}$$

Where $\varphi$ is the dispersion parameter, and $\xi$ is the Tweedie parameter (shape parameter). [14, 16, 17, 28]

Tweedie distributions enjoy many useful properties, including that they are reproductive, where distributions added together to form new distributions are themselves Tweedie. Another is how many of the exponential family distributions converge to Tweedie distributions. [14, 16, 17, 28]

**Common Tweedie Models**

| Family Distribution | Dispersion (extra 0's and tail fatness) | Power (variance proportional to mean: 1/Power) |
|---|---|---|
| Normal | 1 | 0 |
| Poisson | 1 | 1 |
| Compound Poisson | 1 | >1 and <2 |
| Gamma | 1 | 2 |
| Inverse-Gaussian | 1 | 3 |
| Stable | 1 | >2 (Extreme >3) |
| Negative Binomial | >1 | 1 |
| Underdispersion Poisson | <1 | 1 |
| Unique Tweedie | >=1 | >=0 |

Table1: Common Tweedie Models

**The Problem**

However, many problems in industry are not easy to formulate exactly through the exponential family, and linear regression in general has many assumptions that may not be met in scientific and industry data. Machine learning algorithms provide alternative ways to minimize the error between predicted values and actual values of a test set. The sum of squared error is typically used by such optimization algorithms. Many supervised learning algorithms are extensions of generalized linear models and have link functions built into the algorithm to model different outcome distributions; examples are boosted regression, Morse-Smale regression, differential-geometry-based LARS, and Bayesian model averaging. Methods like deep learning and classical neural networks attempt to solve this problem through a series of general mappings leading to a potentially novel link function. [8]

The packages available in Python, R or SAS provide various options for the link function for both GLMs and many of these machine learning models. However, such models struggle when the underlying distribution has a lot of zeros or outliers, and, unfortunately, it is common that none of the link functions provided in these packages is the optimum choice for the given dataset. In addition, the computational cost to solve for the right parameters is very high, particularly for boosted regression, topology-based regression methods, and deep learning models. [8, 9, 14]

**The Problem of Overdispersion in Tweedie Models**

Dispersion parameters relate to the variance of the outcome, and this is well-formulated and modeled explicitly through Tweedie models. Many other GLM links and their machine learning extensions struggle when it comes to the problem of overdispersion, and Tweedie regression itself can struggle when overdispersion is high (including distributions with long tails and distributions with zero-inflation). Simulations in prior papers show this degradation of regression performance, particularly as dispersion parameter increases substantially (values of 4+), including a prior paper that explored bagged KNN regression models as a solution to this problem. Models that worked well on simulated and real datasets

(including the UCI Forest Fire dataset considered in this paper) tended to have long compute times, including the KNN regression ensemble with varying k parameters, which showed the best performance. [8, 14]

## The Solution

Commonly-used generalized linear models with link functions that have been well-explored are among the simplest instances of link-based statistical models, which are based on the underlying geometry of an outcome's underlying probability distribution. However, many others exist, and deep results regarding the exponential family's relation to affine connections in differential geometry provide a possible alternative to link functions. Many more link functions are possible, and by continuously deforming known distributions, one can create new link functions. However, continuous deformations of known distributions will create a very large set of new possible link functions, and algorithms that superpose all possible distributions and collapse to fit a dataset would be ideal to leverage this fact. [1, 21, 26]

Luckily, some quantum computer gates, such as the non-Gaussian transformation gate, essentially perform the first natively and in a computationally-efficient way can be leveraged. Exploiting the geometric relationships between distributions through a superposition of states collapsed to the "ideal" link would present an optimal solution to the problem. [18]

This project provides a proof-of-concept for leveraging specific hardware gates to solve the affine connection problem, including benchmarking of the new algorithm to state-of-the-art levels on simulated and real datasets. Results can be extended to many other, more complicated statistical models, such as generalized estimating equations, hierarchical regression models, and even homotopy-continuation problems. This result is significant and can be a game changer in several disciplines that use GLMs!

**Differential Geometry and the Exponential Family**

It is possible to formulate the exponential family's distributions and parameterizations to form a series of curves on a 2-dimensional surface, where each curve is defined by 2 points at either end of the probability function (0 and 1) and connected by a line that follows a shortest path following parameterization of the distribution, called the geodesic. Since the exponential family can be generalized into Tweedie distributions through continuous transformations, the geodesic connecting 0 and 1 can flow across distributions defining the 2-dimensional surface in a continuous manner (much like homotopy-continuation methods). This yields an affine connection, a morphing of the line as it passes parameters transforms one distribution to another. [1, 21, 26]

Consequently, analytically-derived results/equations for one distribution can be morphed to fit another distribution through continuous transformations! Limit theorems can then be derived by continuous deformations of either moment generating functions or characteristic functions.

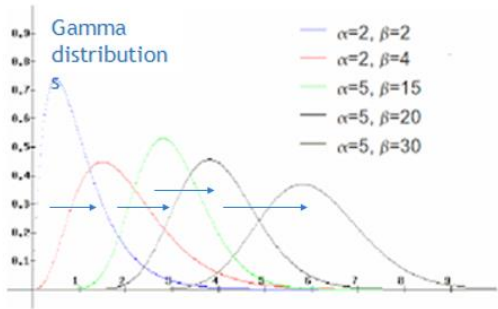

Figure 1: Gamma distribution

## Methodology

The QGLM approach was first tested on a simulated dataset with a high dispersion. And then the approach was tested on a real-life forest fire dataset.

### Simulation

The data simulation was done similar to the approach used in the KNN ensemble paper. 1000 observations were generated, and a 70/30 test/train split was used. 1 noise variable was also added. [8] The following parameters were used in the simulation

                Tweedie parameter=1

                dispersion parameter=8

### Empirical dataset

UCI Repository's Forest Fire dataset was chosen for testing this approach since it is hard to model and several machine learning models fail to beat the mean model. The data has 12 predictors (2 spatial coordinates of location, month, day, FFMC index, DMC index, DC index, ISI index, temperature, relative humidity, wind, and rain) and 517 observations. [8]

t-SNE was used to reduce the dimensionality of the predictor set to 4 components to make it compatible with Xanadu's capabilities. Once again, a 70/30 test/train split was used. [29, 30]

### Comparison methods

The following algorithms/models were also applied to the simulation data as well as the UCI Forest Fires dataset.

1. Boosted regression with linear baselearners (R package mboost) [13]
2. Random forest, a tree-based bagged ensemble (R package ranger) [3]
3. DGLARS, tangent-space-based least angle regression model (R package dgLARS) [2]
4. BART, a Bayesian-based tree ensemble (R package BART) [22]
5. HLASSO, a homotopy-based LASSO model (R package lasso2) [25]
6. Poisson regression (R GLM function without any modifications)

The performance of the various algorithms was then compared. The outcomes are presented in the results section of this paper.

# Technology

## Xanadu and Suitability to GLM

There are several proposed architectures for building quantum computers. One of the architectures is based on the continuous variable (CV) model. The Xanadu implementation of a CV based quantum computing is based on qumodes, which are the basic computational units of the architecture. [18]

The continuous representation used in this model is very well-suited for modelling complex probability distributions, and the non-Gaussian transformation gates available in this implementation provide perfect avenue to perform the affine transformation related to the outcome distribution without a need to specific a link function to approximate the geometry. [18]

This formulation has the potential to approximate any continuous outcome's distribution, creating potential new "link functions" through this gate through affine transformation of the wavefunctions representing the data. This removes the need for approximations by easy-to-compute link transformations. In addition, it can approximate any continuous distribution, including ones that aren't included in common statistical packages implementing GLMs and their longitudinal/survival data extensions. Thus, Xanadu's system provides a general solution to the linear regression equation with many potential extensions to more sophisticated regression models!

Moreover, GLMs and their extensions are all based on simple matrix operations. Matrix multiplication and addition for a linear model coupled with a continuous transformation of the model results to fit the outcome distribution. Xanadu's qumode formulation makes it ideal for implementing quantum GLMs (QGLMs).

# Data Preprocessing

The dimensionality of the dataset was reduced through t-SNE to create a set of 4 predictors and 1 outcome, such that predictors are uncorrelated when entered into models. Decorrelation helps most regression methods, including linear models and tree models. Other dimensionality reduction methods are possible, including the introduction of factors from factor analytic models or combinations of linear/nonlinear, global/local dimensionality reduction algorithms. [29, 30]

The outcome was scaled to the range -3 to 3, such that the Xanadu simulation can effectively model and process the data in qumodes. A slight warping of the most extreme values, but these are generally less than 5 observations per dataset. In future work, it may be useful to explore other types of scaling.

**Qumodes Circuit Details**

GLMs can be embedded within Xanadu's qumode quantum computer simulation software (and qumode computer) with a singular value decomposition of the $\beta$ coefficient in the formulation [18]:

$$Mean(Y) = g^{-1}(X\beta)$$

This translates to $\beta = O_1 \Sigma O_2$, which can be modeled through a series of quantum circuit gates:

    a. Multiplication of X and an orthogonal matrix:

$$|O_1 X\rangle \cong U_1 |X\rangle$$

which corresponds to a linear interferometer gate ($U_1$) acting on X

    b. Multiplication of that result by a diagonal matrix:

$$|\Sigma O_1 X\rangle \propto S(r)|O_1 X\rangle$$

which corresponds to a squeezing gate that acts on a single qumode

c. Multiplication of X and an orthogonal matrix:

$$|O_2 \Sigma O_1 X\rangle \cong U_2|\Sigma O_1 X\rangle$$

which corresponds to a linear interferometer gate ($U_2$) acting on the result

d. Multiplication by a nonlinear function on this result:

$$|g^{-1}(O_2 \Sigma O_1 X)\rangle \cong \Phi|O_2 \Sigma O_1 X\rangle$$

which corresponds to the non-Gaussian gate acting on the result

This gives a final result of gates acting upon the dataset as:

$$\Phi * \mathcal{U}_2 * \mathcal{S} * \mathcal{U}_1 |X\rangle \propto |g^{-1}(X\beta)\rangle$$

**Strawberry Fields**

Strawberry Fields is a software architecture from Xanadu for simulating a photonic quantum computer's circuitry and running algorithms on those circuits. It is a full-stack software platform and it is open source, with implementation in Python. This platform is meant for implementation of the CV model of quantum computing. It implements a quantum programming language called 'Blackbird.' Strawberry Fields also offers three simulated backends which are implemented in numpy and tensorflow. [19]

**Qumodes Parameter Settings**

The deep learning framework was already available in the implementation. For our experiment, the hidden layers and bias terms were removed to collapse the algorithm to a generalized linear model framework. The loss function optimized was mean square error, which corresponds to the loss functions specified in the comparison algorithms. The qumode cut-off dimension was set to 10. Optimization via least squares was not available, so gradient descent was used with a learning rate of 0.1 over 80 iterations, giving a qumodes implementation of a quantum generalized linear model with a boosting flavor to it. Because the quantum computing component is inherently probabilitistic, algorithms were run on the same training and test set 10 times apiece to average out quantum effects. [18]

# Results

Results for the simulated dataset were encouraging, and the overdispersion can be easily seen in the histogram of the training dataset:

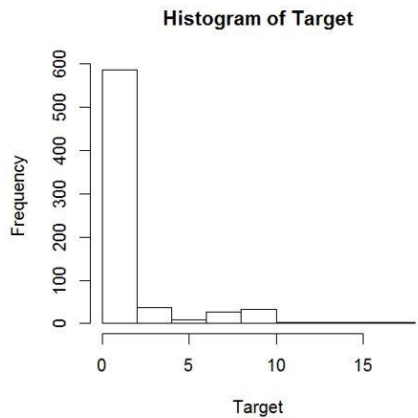

Figure 2: Histogram of the target

| Algorithm | Scaled Model MSE |
| --- | --- |
| Random Forest | 0.80 |
| BART | **0.78** |
| Boosted Regression | **0.78** |
| DGLARS | 0.81 |
| HLASSO | 0.81 |
| GLM | 0.81 |
| QGLM | 0.82 |
| Mean | 0.85 |

Table 2: Error values for various algorithms when applied to the simulated dataset

QGLMs yield slightly worse prediction on the simulated dataset. However, their performance is not far off from state-of-the art algorithms, and some random error is expected from the quantum machinery. This suggests that they are a reasonable alternative to well-studied algorithms like boosted regression and dgLARS.

Examining the performance on UCI's Forest Fire dataset, we see that QGLMs emerge as the best-performing algorithm on this difficult, real-world dataset:

| Algorithm | Scaled Model MSE |
| --- | --- |
| Random Forest | 0.125 |
| BART | 0.125 |
| Boosted Regression | 0.119 |
| DGLARS | 0.114 |
| HLASSO | 0.120 |
| GLM | 0.119 |
| **QGLM** | **0.106** |
| Mean | 0.115 |

Table 3: Error values for various algorithms when applied to the UCI Forest Fires dataset

In fact, QGLMs provide ~10% gain over the next best algorithm on this dataset. This suggests that they work well on real data and difficult problems.

## Conclusion

The qumodes formulation with its unique operators can eliminate the need for link functions within linear models by exploiting the exponential family's underlying geometry and still give good prediction on difficult regression problems. In this paper, QGLMs based on this notion achieved better than state-of-the-art prediction for a difficult Tweedie regression dataset (UCI Forest Fire) and produced comparable results on a difficult simulation dataset.

This result has the potential to bring statistical modeling and quantum computing together, by leveraging the similarity in the underlying geometry of the dataset and the quantum gate. The potential future areas of application and research include generalized estimating equations, generalized linear mixed models, structural equation models and hierarchical regression models. In addition, this offers a potential avenue through which to implement the homotopy-continuation method, which is common in dynamic systems modeling.